\pdfoutput=1
\documentclass[aps, prb, superscriptaddress, twocolumn, dvipdfmx]{revtex4-1}
\usepackage{siunitx}
\usepackage[pdftex]{graphicx}
\usepackage{amsmath}
\newcommand{\Tc}{$T_{\mathrm{c}}$}
\begin{document}
\title{Charge carrier dynamics of FeSe thin film investigated by\\
terahertz magneto-optical spectroscopy}

\author{Naotaka Yoshikawa}
\affiliation{Department of Physics, The University of Tokyo, Hongo, Tokyo 113-0033, Japan}

\author{Masayuki Takayama}
\affiliation{Department of Physics, The University of Tokyo, Hongo, Tokyo 113-0033, Japan}

\author{Naoki Shikama}
\affiliation{Department of Basic Science, The University of Tokyo, Meguro, Tokyo 153-8902, Japan}

\author{Tomoya Ishikawa}
\affiliation{Department of Basic Science, The University of Tokyo, Meguro, Tokyo 153-8902, Japan}

\author{Fuyuki Nabeshima}
\affiliation{Department of Basic Science, The University of Tokyo, Meguro, Tokyo 153-8902, Japan}

\author{Atsutaka Maeda}
\affiliation{Department of Basic Science, The University of Tokyo, Meguro, Tokyo 153-8902, Japan}

\author{Ryo Shimano}
\affiliation{Department of Physics, The University of Tokyo, Hongo, Tokyo 113-0033, Japan}
\affiliation{Cryogenic Research Center, The University of Tokyo, Yayoi, Tokyo, 113-0032, Japan}

\begin{abstract}
We performed terahertz magneto-optical spectroscopy of FeSe thin film to elucidate the charge carrier dynamics. The measured diagonal (longitudinal) and off-diagonal (Hall) conductivity spectra are well reproduced by two-carrier Drude model, from which the carrier densities, scattering times and effective masses of electron and hole carriers are determined in a wide range of temperature. The hole density decreases below the structural transition temperature while electron density increases, which is attributed to the band structure modification in the electronic nematic phase. The scattering time of the hole carrier becomes substantially longer than that of the electron at lower temperature, which accounts for the increase of the positive dc Hall coefficient at low temperature.
\end{abstract}
\maketitle

Since the discovery of iron-based superconductors (FeSCs), tremendous research efforts have been devoted to reveal the pairing mechanism of superconductivity. The elucidation of interplay between the nematic order, antiferromagnetic spin order, and superconductivity emergent in FeSCs has been believed to provide a clue to understand the emergent superconductivity. Among FeSCs, FeSe provides a unique playground to study the role of nematicity, because it lacks the long-range magnetic order in the nematic phase that appears below the tetragonal-orthorhombic structural transition temperature  $T_{\mathrm{s}}\simeq \SI{90}{\kelvin}$, as evidenced by a significant electronic anisotropy from transport and nuclear magnetic resonance (NMR) spectral properties\cite{McQueen:2009hs,Baek:2014gs,Bohmer:2015fk}. While the superconducting transition temperature $T_{\mathrm{c}}$ of bulk FeSe is $\sim\SI{9}{K}$ at ambient pressure\cite{Hsu:2008ep}, it shows a remarkable tunability. $T_{\mathrm{c}}$ increases to as high as \SI{38}{K} under hydrostatic pressure\cite{Medvedev:2009ex,Imai:2009hw,Mizuguchi:2008bn,Sun:2016dh}, and single-layer FeSe grown on SrTiO$_3$ shows $T_{\mathrm{c}}$ up to \SI{109}{K}\cite{Ge:2014hc,He:2013cn,Tan:2013jb}. Electron doping by ionic-gating in FeSe thin flakes enhances the superconductivity toward \SI{48}{K}\cite{Lei:2016gl,Shiogai:2015fw,Hanzawa:2017fa,Kouno:2018fp}. Intercalation also enhances \Tc{} by a similar doping effect in addition to an effect of separating the layers\cite{BurrardLucas:2012fm}. One important key to understand the \Tc{} increase of FeSe is considered to be a change of the Fermi surface topology. The high tunability of the electronic structure of FeSe achieved by various ways is related to its extremely small effective Fermi energy, which has been demonstrated in FeSe\cite{Kasahara:2014gt} as well as FeSe$_{1-x}$Te$_x$\cite{Lubashevsky:2012br,Okazaki:2014im}. FeSe is a semimetal with the Fermi surface consisting of hole pockets around the Brillouin zone center $\mathit{\Gamma}$ point and electron pockets around the zone corner $M$ point. The low-energy electronic structure around the Fermi level has been experimentally revealed by angle-resolved photoemission spectroscopy (ARPES)\cite{Shimojima:2014kc,Zhang:2015fx,Watson:2015kn,Nakayama:2014eo,Fanfarillo:2016kz}. ARPES studies have also shown a significant modification of the band structure below $T_{\mathrm{s}}$ which is attributed to the development of an electronic nematicity.

For the understanding of unconventional superconductivity in FeSe, it is also indispensable to investigate the charge carrier dynamics in normal and nematic phases as well as in superconducting phase. The Hall resistivity measured by magneto-transport shows an unusual temperature dependence with the sign change owing to the nearly compensated electron and hole carriers\cite{Kasahara:2014gt,Watson:2015hx,Sun:2016by,Nabeshima:2018fi}. In bulk FeSe, the presence of a small number of highly mobile electron-like carrier at the nematic phase was also identified by the Hall resistivity\cite{Watson:2015hx} and mobility spectrum analysis\cite{Huynh:2014ch}, which could be attributed to the Dirac-like dispersion near the $M$ point\cite{Tan:2016cd}. However, the complexity of the multi-band Fermi surfaces of FeSe makes it difficult to grasp the properties of charge carriers only by dc transport measurements. This is because the characterization of the carriers by dc transport measurements needs to assume some models such as a compensated two-band model, where the compensated electron is assumed to have same carrier density as that of the hole ($n_e=n_h$)\cite{Watson:2015hx,Nabeshima:2018fi,Huynh:2014ch,Ovchenkov:2017fp}. Although three band model can also be used by including the nonlinear term when dc transport is measured up to high magnetic field, the characterization is not complete because the mobility, which is determined by dc transport in addition to carrier densities, is a function of effective mass and scattering time. For more detailed characterization of charge carriers, quantum oscillations are a well-established technique\cite{Watson:2015kn,Watson:2015hx,Terashima:2014ft,Audouard:2015hp}. However, quantum oscillations are able to be observed typically in bulk crystals grown by the vapor transport techniques. Thus, the observation of quantum oscillations of FeSe has been limited in bulk single crystals and at very low temperature. Therefore, the properties of charge carriers of FeSe in a wide range of temperature in particular across the structural phase transition temperature have remained to be clarified.

In this study, we investigate the charge dynamics in a thin-film FeSe by terahertz (THz) magneto-optical spectroscopy. The obtained diagonal (longitudinal) and off-diagonal (Hall) conductivity spectra are well described by two-carrier Drude model, from which the carrier densities, scattering times and effective masses of electron and hole carriers were independently determined. The temperature dependence of THz magneto-optical spectra revealed the significant change of the carrier densities below $T_{\mathrm{s}}$, which is plausibly attributed to the band structure modification in the nematic phase. The scattering time of the hole carrier substantially increases at lower temperature, which explains the peculiar temperature dependence of the dc Hall coefficient in FeSe thin films.

\begin{figure}
\includegraphics[width=\columnwidth]{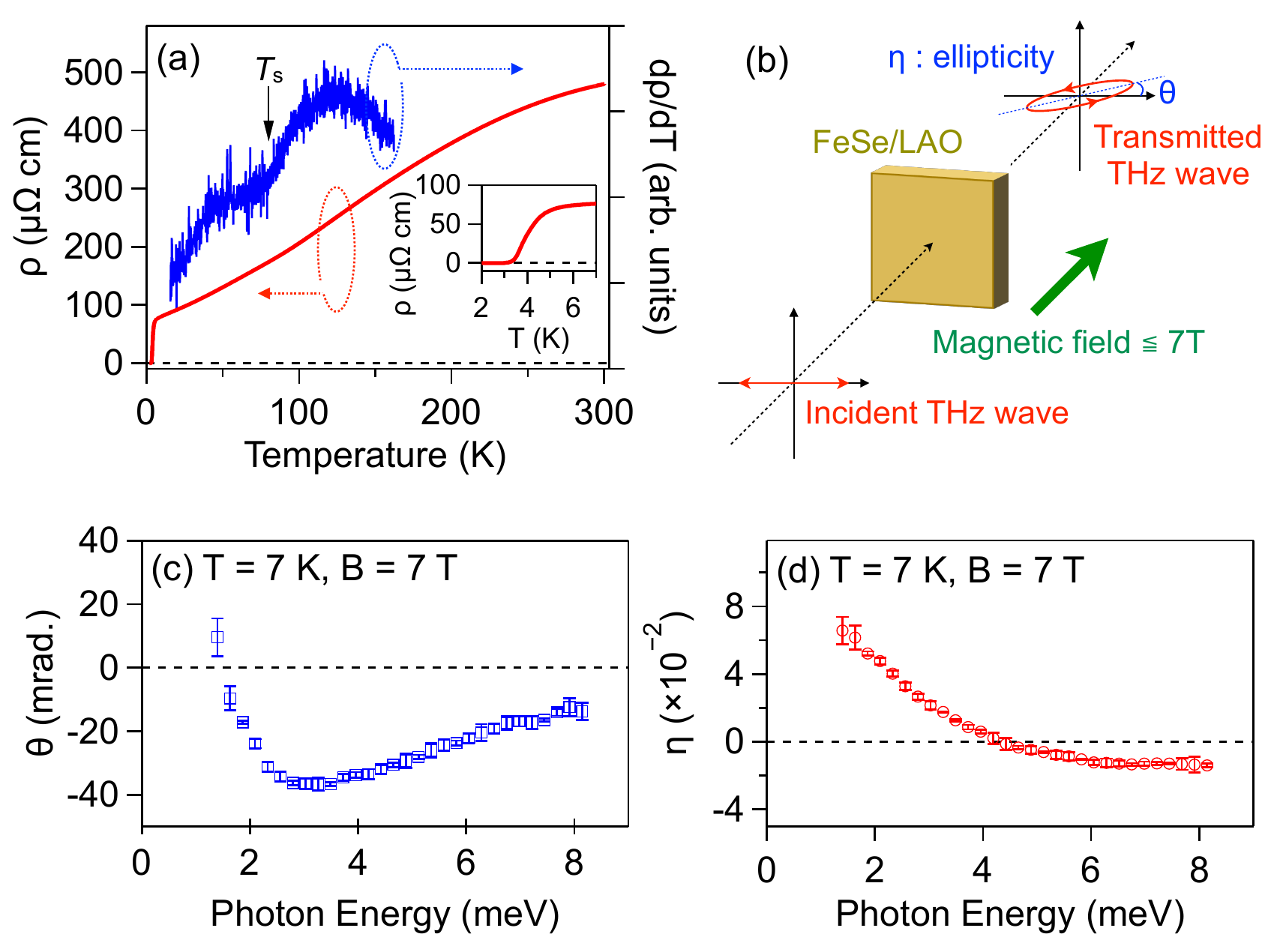}
\caption{(a)~Temperature dependence of dc resistivity $\rho$ (red line) and $d\rho/dT$ (blue line) of the FeSe thin film. A kink anomaly at $T_{\mathrm{s}}$ is indicated by black arrow. Inset shows an enlarged view of the resistivity curve around \Tc{}~$\sim\SI{3}{K}$. (b)~Schematic of our THz magneto-spectroscopy. (c)~Faraday rotation spectrum and (d)~ellipticity spectrum induced by FeSe film at \SI{7}{K} with the magnetic field of \SI{7}{T}.}
\end{figure}

A FeSe thin film with the thickness of 46 nm was fabricated on LaAlO$_3$ (LAO) substrate by pulsed-laser deposition method\cite{Imai:2010ez,Imai:2010jl}. The temperature dependence of dc resistivity shows the superconducting transition at \Tc{}~$\sim\SI{3}{K}$ defined by the zero resistivity (Fig.~1(a)). A kink anomaly in $d\rho/dT$ curve indicates the structural transition at $T_{\mathrm{s}}\sim \SI{80}{K}$. Figure~1(b) shows the schematic of our THz magneto-spectroscopy based on THz time-domain spectroscopy (THz-TDS)\cite{Ikebe:2008cq,Ikebe:2009it,Shimano:2013ez}. The output of a mode-locked Ti:sapphire laser with the pulse duration of 110 fs, center wavelength of 800 nm, and repetition rate of 76 MHz was focused onto a p-type $(111)$ InAs surface to generate THz pulses. Linearly polarized THz incident pulses were focused on the sample placed in a split-type superconducting magnet which can produce the magnetic field up to \SI{7}{T} in Faraday configuration, that is, the magnetic field is parallel to the wavevector of the THz wave. The THz-wave was detected by electro-optical sampling with a $(110)$ ZnTe crystal. By measuring the waveform of the parallel polarization component defined as $E_x(t)$ and perpendicular polarization component $E_y (t)$ of the transmitted THz pulses by using wire-grid polarizers, the Faraday rotation angle $\theta$ and ellipticity $\eta$ induced by the FeSe film in the magnetic field can be obtained. Here, the approximated expression $E_y (\omega)/E_x (\omega)\sim \theta(\omega)+i\eta(\omega)$ for small Faraday rotation angle was used. Figures~1(c) and 1(d) show the THz Faraday rotation angle and ellipticity, respectively, under $B = \SI{7}{T}$ at the sample temperature $T = \SI{7}{K}$. The error bars indicate the standard deviations determined by the multiple measurements, confirming that the obtained signal $\theta(\omega)$ and $\eta(\omega)$ well exceed the noise level.

Figures 2(a) and 2(b) show the real- and imaginary-part of the longitudinal optical conductivity spectrum $\sigma_{xx} (\omega)$ of FeSe at \SI{7}{K}, respectively, given by the normal transmission type THz-TDS without magnetic field. The combination of the obtained $\sigma_{xx} (\omega)$, $\theta(\omega)$ and $\eta(\omega)$ provides the optical Hall conductivity spectrum $\sigma_{xy} (\omega)$ through the following equation for a thin film under small rotation angle approximations:
\begin{align}
\theta+i\eta\sim\frac{\sigma_{xy}(\omega)d}{\left( 1+n_{\mathrm{sub}}(\omega)-\frac{i\omega dn_{\mathrm{sub}}(\omega)}{c}\right)c\varepsilon_0 + \sigma_{xx}(\omega)d}
\end{align}
where $n_{\mathrm{sub}}$  is the refractive index of the substrate, $d$ the thickness of FeSe film, $c$ speed of light, and $\varepsilon_0$ permittivity of vacuum. Figures 2(c) and 2(d) show the real- and imaginary-part of the optical Hall conductivity spectrum with $B = \SI{7}{T}$, respectively. The data set of the complex longitudinal and Hall conductivity spectra was well fitted by the Drude model with an electron carrier and a hole carrier:
\begin{align}
\sigma_{xx}(\omega,B=0)&=\frac{n_e q_e^2 \tau_e}{m^*_e}\frac{1}{1-i\omega \tau_e}+\frac{n_h q_h^2 \tau_h}{m^*_h}\frac{1}{1-i\omega \tau_h}\\
\sigma_{xy}(\omega,B)&=\frac{n_e q_e^2 \tau_e}{m^*_e}\frac{\omega_{c,e}\tau_e}{(1-i\omega \tau_e)^2-\omega_{c,e}^2}\notag\\
&\quad+\frac{n_h q_h^2 \tau_h}{m^*_h}\frac{\omega_{c,h}\tau_h}{(1-i\omega \tau_h)^2-\omega_{c,h}^2}
\end{align}
\begin{figure}
\includegraphics[width=\columnwidth]{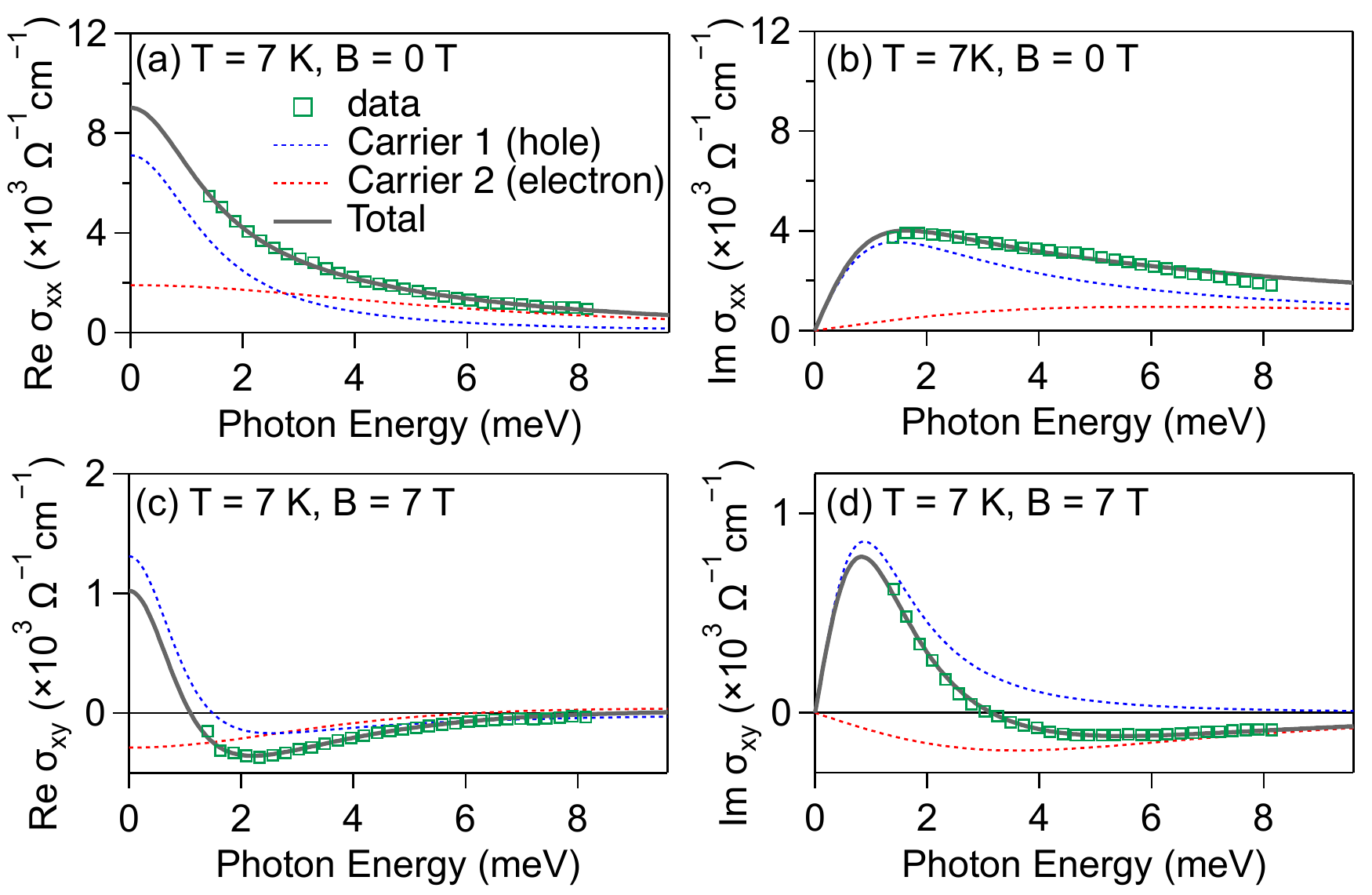}
\caption{(a) Real- and (b) imaginary-part of the longitudinal optical conductivity spectrum, and (c) real- and (d) imaginary-part of the optical Hall conductivity spectrum, respectively, of FeSe at $T=\SI{7}{K}$ with $B=\SI{7}{T}$. Green open square represents experimental result, blue dashed, red dashed, and grey solid line indicate the contribution of the hole carrier, electron carrier, and sum of them given by the two-carrier Drude fitting described in the text.}
\end{figure}
where $q_e=-e$ ($q_h=+e$), $n_e$ ($n_h$), $\tau_e$ ($\tau_h$), $m_e^*$ ($m_h^*$), $\omega_{c,e}=q_e B/m_e^*$ ($\omega_{c,h}=q_h B/m_h^*$) represent the charge of carrier, carrier density, scattering time, effective mass, cyclotron frequency of the electrons (holes), respectively. Notably, the optical Hall conductivity is sensitive to the carrier type (electron-like or hole-like) while the longitudinal conductivity is formally independent of the sign of the charge carriers. The sign change around \SI{3}{meV} in the imaginary-part of $\sigma_{xy} (\omega)$ indicates that two types of carriers exist with different sign of charge, which can be attributed to those in the electron pocket and hole pocket in FeSe. We can determine the complete set of the parameters describing charge carrier dynamics, $n_e=\SI{5.4E19}{cm^{-3}}$, $n_h=\SI{1.7E20}{cm^{-3}}$, $m_e=0.86m_0$, $m_h=2.90m_0$, $\tau_e=\SI{0.11}{ps}$, $\tau_h=\SI{0.45}{ps}$. We also checked the magnetic field dependence of the complex optical Hall conductivity. The good agreement between the experiments and the two-carrier Drude model is also confirmed in the magnetic field dependence of the diagonal and off-diagonal conductivity spectra. The effective mass of the hole evaluated by quantum oscillation and ARPES studies are typically around $4m_0$, which is consistent with our result\cite{Watson:2015kn,Terashima:2014ft,Audouard:2015hp,Phan:2017ew}. The effective mass of the electron, on the other hand, differs among various reports. The present THz magneto-spectroscopy shows the similar effective mass with that deduced by the band curvature measured by ARPES\cite{Phan:2017ew,Watson:2016dy}. The carrier densities $n_e$ and $n_h$ are also reasonable compared with the various transport measurements\cite{Watson:2015hx,Nabeshima:2018fi,Huynh:2014ch}. The ARPES measurements showed that the Fermi surface around the zone center consists of the outer and inner hole pockets above the structural transition temperature, while the inner hole pocket moves completely below the Fermi level in the nematic phase\cite{Watson:2015kn}. Thus, the carrier density and effective mass of the hole at \SI{7}{K} evaluated by the THz magneto-spectroscopy are most likely attributed to that of the outer hole pocket. On the other hand, the Fermi surface around $M$ point consists of at least two electron pockets in the nematic phase. Here the electrons in the outer electron pocket are considered to contribute dominantly to the transport and charge dynamics, so the obtained parameters of the electron are probably those of the outer electron pocket.

\begin{figure}
\includegraphics[width=\columnwidth]{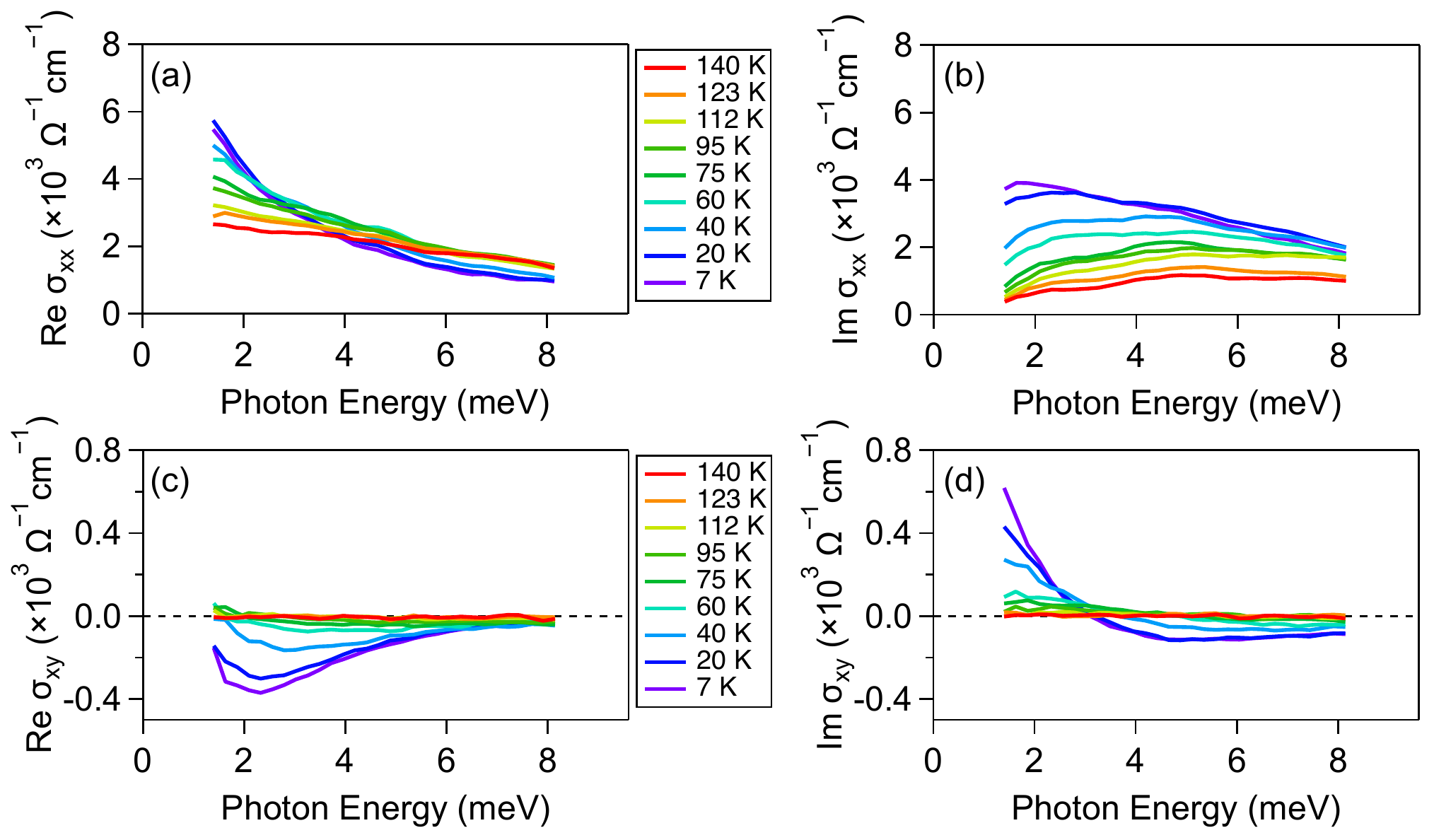}
\caption{Temperature dependence of (a)~real- and (b)~imaginary-part of the longitudinal optical conductivity, and (c)~real- and (d)~imaginary-part of the optical Hall conductivity of FeSe thin film with $B=\SI{7}{T}$.}
\end{figure}
Next, we investigated the temperature dependence of conductivity spectra. Figures 3(a) and 3(b) show the complex longitudinal optical conductivity spectra from \SI{7}{K} to \SI{140}{K}. The conductivity spectrum shows a broadening at higher temperature indicating the shorter scattering time of the carriers. The optical Hall conductivity spectra in Figs.~3(c) and 3(d) shows more notable change where it goes to almost zero at \SI{140}{K}. We performed the two-carrier Drude fitting as done in Fig.~2 for various temperature and extracted the parameters describing the observed complex conductivity spectra. Here we fixed effective masses of the electron and hole to those obtained for \SI{7}{K}. From the fitting, we elucidate that zero Hall conductivity in the THz frequency at \SI{140}{K} is a consequence of almost completely compensated electron and hole carriers at this temperature. Figure~4 summarizes the temperature dependence of the extracted charge carrier parameters. The scattering time of the hole is shortened at higher temperature while that of the electron weakly depends on the temperature as shown in Fig.~4(a). By considering that the impurity scattering dominates at $T=0$ limit, it is found that the impurity scattering rate of the electron is larger than that of the hole. This could be derived from the about three times lighter effective mass of the electron than the hole. Since the ARPES measurements revealed that the Fermi wave numbers $k_F$ of the electron and the hole pockets are similar to each other\cite{Shimojima:2014kc,Zhang:2015fx,Watson:2015kn,Phan:2017ew,Watson:2016dy}, the Fermi velocity $v_{\mathrm{F}}=k_{\mathrm{F}}/m^*$ of the electron is larger than that of the hole. The carrier with the larger Fermi velocity is expected to have the larger scattering rate with the impurities with certain density. Concomitantly, the hole density substantially reduces at lower temperature below $T_{\mathrm{s}}$ while that of electrons increases. These behaviors seem to correlate with the band structure change between tetragonal and orthorhombic phase accompanied by electronic nematicity. As for the hole pockets around $\mathit{\gamma}$ point, the ARPES revealed that the inner hole band moves below the Fermi energy in the nematic phase\cite{Watson:2015kn}, leading to the smaller hole carrier density. The $d_{xy}$ band which forms the outer electron-pocket at the $M$ point is also pushed down with decreasing temperature, resulting in the increase of electron density at lower temperature below $T_{\mathrm{s}}$. A decrease of carrier densities below the structural transition temperature has been observed in bulk FeSe by the dc magneto-transport measurement\cite{Watson:2015hx}. The strong anisotropy of the scattering rate induced by the enhancement of spin fluctuations below $T_{\mathrm{s}}$ was suggested to explain the drop of the effective electron and hole densities without a change of Fermi surface volume. However, our observation of the simultaneous increase of the hole density and decrease of the electron density cannot be explained only by the strongly anisotropic carrier scattering. The present THz magneto-optical spectroscopy unambiguously reveals the temperature dependence of the electron and hole densities, which suggests the Fermi surface modification in the nematic phase. The reduction of carrier density below $T_{\mathrm{s}}$ has also been observed by conventional far-infrared optical reflectivity measurement\cite{Nakajima:2017cw}. By using the scattering time and effective mass, we further evaluate the mobility $\mu=e\tau/m^*$ for each carrier as shown in Fig.~4(c). The mobility of the hole increases at lower temperature as a result of the increase of the scattering time. The temperature dependence of the dc Hall coefficient described as $R_{\mathrm{H}}\approx \sigma_{xy}(\omega=0)/(\sigma_{xx}(\omega=0)^2 B)$ is plotted (red circles) in Fig. 4(d). It shows an excellent agreement with the Hall coefficient obtained by dc magneto-transport measurement (grey solid curve), indicating that the dc transport and the THz response are described by the common physical origin of the charge carrier dynamics. Accordingly, the peculiar temperature dependence of the Hall coefficient that increases monotonically toward the low temperature is dominantly attributed to the increase of hole scattering time.
\begin{figure}
\includegraphics[width=\columnwidth]{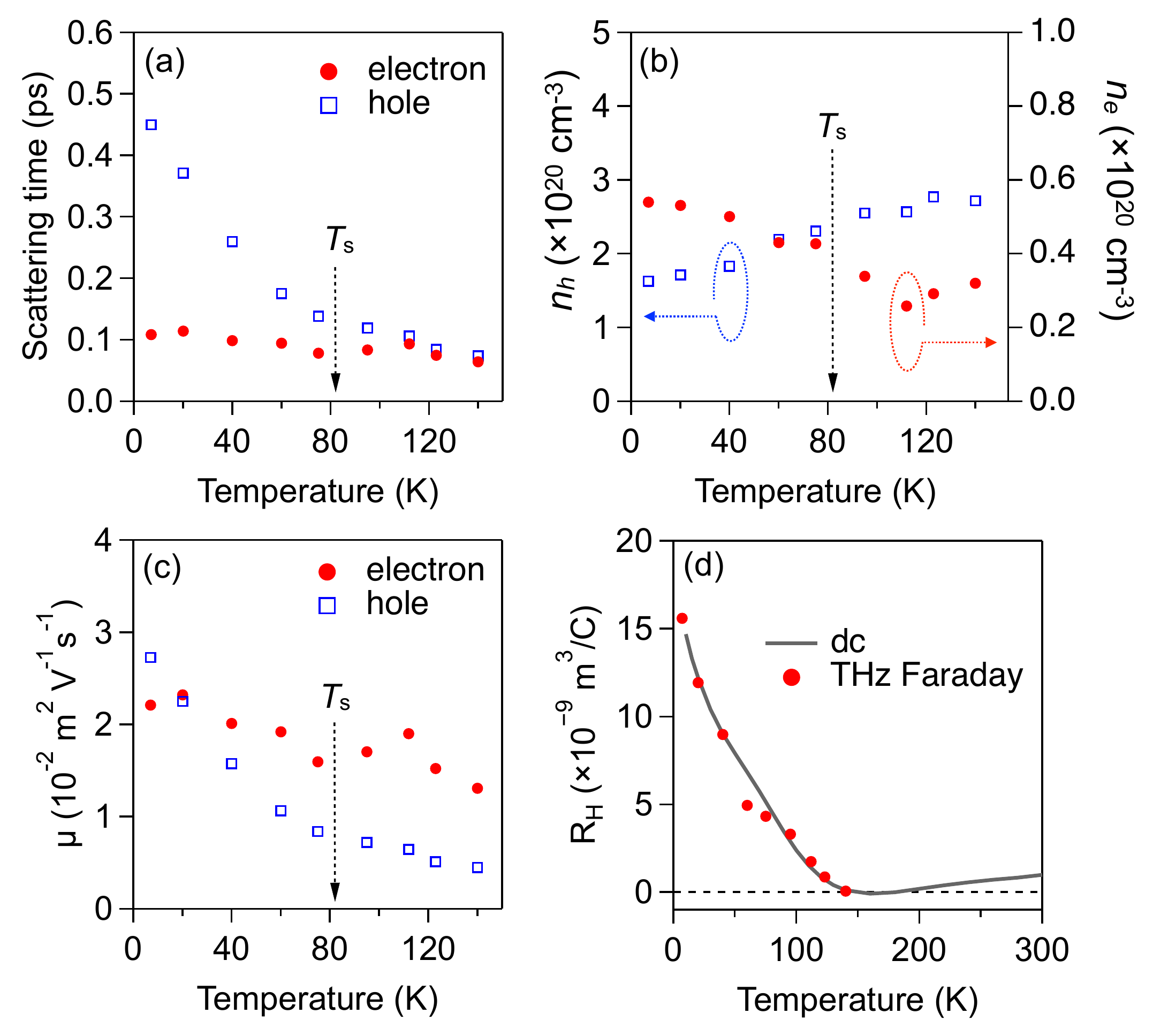}
\caption{Temperature dependence of (a) scattering times, (b) carrier densities, (c) mobilities of the electron and hole of FeSe thin film given by THz magneto-spectroscopy. The structural transition temperature Ts is indicated by arrows. (d) Hall coefficient of FeSe thin film as a function of temperature, obtained by THz magneto-spectroscopy (red circles) and dc transport measurement (grey line).}
\end{figure}

In summary, we performed THz magneto-optical spectroscopy to investigate the charge carrier dynamics in FeSe. The obtained diagonal and off-diagonal conductivity spectra are well described by two-carrier Drude model, from which the carrier densities, scattering times and effective masses of electron and hole carriers are determined in a wide temperature range. We found the significant temperature dependence of the carrier densities of electrons and holes below $T_{\mathrm{s}}$ , which is most likely attributed to the band structure modification at the structural phase transition. The scattering time of the hole carrier becomes substantially longer than that of the electron at lower temperature, which results in the positive dc Hall coefficient at low temperature observed in the present FeSe thin film. We demonstrated that THz magneto-optical spectroscopy is a powerful tool for investigating the charge carrier dynamics of FeSe. The application of the technique to FeSe system under pressure, with ionic gating, and thin film or single-layer on various substrates would pave a unique pathway to access the charge carrier properties, thereby providing a deep insight into their correlation with the nematicity and the superconductivity.

This work was supported by JSPS KAKENHI (Grants No. 18H05324, No. 18H05846, and 18H04212).

\bibliographystyle{apsrev4-1}
\bibliography{refs_FeSeFaraday}

\end{document}